\documentclass[
twocolumn,prd,
showpacs,
nofootinbib,
amsmath,amssymb,
superscriptaddress]{revtex4-1}

\usepackage{float}
\usepackage{graphicx}% Include figure files
\usepackage{dcolumn}% Align table columns on decimal point
\usepackage{bm}% bold math
\usepackage{hyperref}% add hypertext capabilities
\usepackage{ulem} %xout command
\usepackage{subcaption}
\usepackage{gensymb}
\newcommand{\be}{\begin{equation}}
\newcommand{\ee}{\end{equation}}
\newcommand{\bi}{\begin{itemize}}
\newcommand{\ei}{\end{itemize}}
\newcommand{\bea}{\begin{eqnarray}}
\newcommand{\eea}{\end{eqnarray}}

\def\gw#1{gravitational wave#1 (GW#1)\gdef\gw{GW}}
\def\snr#1{signal-to-noise ratio#1 (SNR#1)\gdef\snr{SNR}}
\def\bh#1{black hole#1 (BH#1)\gdef\bh{BH}}
\def\bbh#1{binary black hole#1  (BBH#1)\gdef\bbh{BBH}}
\def\bns#1{binary neutron star#1 (BNS#1)\gdef\bns{BHS}}
\def\nr#1{numerical relativity#1 (NR#1)\gdef\nr{NR}}
\def\Nr#1{Numerical relativity#1 (NR#1)\gdef\Nr{NR}}
\def\gr#1{general relativity#1 (GR#1)\gdef\gr{GR}}
\def\qnm#1{quasi-normal mode#1  (QNM#1)\gdef\qnm{QNM}}
\def\psd#1{power spectral density#1  (PSD#1)\gdef\psd{PSD}}
\def\fpeak#1{the instantaneous frequency at maximum amplitude#1  ($\omega_{peak}$#1)\gdef\fpeak{$\omega_{peak}$}}
\def\fdotpeak#1{the derivative of the instantaneous frequency at maximum amplitude#1  ($\dot{\omega}_{peak}$#1)\gdef\fdotpeak{$\dot{\omega}_{peak}$}}

\def\fpeaknr#1{the dimensionless instantaneous frequency at maximum amplitude#1  ($\hat{\omega}_{peak}$#1)\gdef\fpeaknr{$\hat{\omega}_{peak}$}}

\def\fdotpeaknr#1{the derivative of the dimensionless instantaneous frequency at maximum amplitude#1  ($\hat{\dot{\omega}}_{peak}$#1)\gdef\fdotpeaknr{$\hat{\dot{\omega}}_{peak}$}}

\def\cm#1{the chirp mass#1 ($\mathcal{M}$#1)\gdef\cm{$\mathcal{M}$}}
\def\af#1{the dimensionless remnant spin#1 ($a_{f}$#1)\gdef\af{$a_{f}$}}

\def\qnmfreq#1{frequency#1  ($\omega_{qnm}$#1)\gdef\qnmfreq{$\omega_{qnm}$}}
\def\qnmdecay#1{decay time#1  ($\tau_{qnm}$#1)\gdef\qnmdecay{$\tau_{qnm}$}}

\def\cwb#1{Coherent WaveBurst#1 (cWB#1)\gdef\cwb{cWB}}

\def\ligo#1{Laser Interferometer Gravitational-wave Observatory#1 (LIGO#1)\gdef\ligo{LIGO}}
\def\virgo#1{Virgo#1\gdef\virgo{Virgo}}
\def\lisa#1{Laser Interferometer Space Antenna#1 (LISA#1)\gdef\lisa{LISA}}
\def\et#1{Einstein Telescope#1 (ET#1)\gdef\et{ET}}
\def\ce#1{Cosmic Explorer#1 (CE#1)\gdef\ce{CE}}
\def\lvc#1{LIGO and Virgo Collaboration#1 (LVC#1)\gdef\lvc{LVC}}
\def\lvk#1{LIGO-Virgo-KAGRA (LVK#1) Collaboration#1\gdef\lvk{LVK}}

\def\etk#1{Einstein Toolkit#1 (ETK#1)\gdef\etk{ETK}}

\def\pn#1{Post Newtonian#1 (PN#1)\gdef\pn{PN}}

\def\massratio#1{mass ratio#1 ($q = m_1 / m_2 > 1$#1)\gdef\massratio{$q$}}

\def\lalsim{LALSimulation}
\def\pycbc{PyCBC}

\usepackage{color}

\usepackage{soul}

\begin{document}

\title{Optimizing the Placement of Numerical Relativity Simulations using a Mismatch Predicting Neural Network}

\affiliation{Center for Gravitational Physics and Department of Physics, The University of Texas at Austin, Austin, TX 78712}

\author{Deborah Ferguson$^{1}$}\noaffiliation
%\preprint{LIGO-xxxxx}
%\pacs{04.80.Nn, 04.25.dg, 04.25.D-, 04.30.-w} 

\begin{abstract}
Gravitational wave observations from merging compact objects are becoming commonplace, and as detectors improve and gravitational wave sources become more varied, it is increasingly important to have dense and expansive template banks of predicted gravitational waveforms.
Since numerical relativity is the only way to fully solve the non-linear merger regime of general relativity for comparably massed systems,  numerical relativity simulations are critical for gravitational wave detection and analysis.
These simulations are computationally expensive, with each simulation placing one point within the high dimensional parameter space of binary black hole coalescences.
%Given the high dimensional parameter space and discrete nature of these simulations, they are often used to build models which can produce arbitrarily dense template banks for data analysis.
%However, creating reliable and accurate models still requires sufficient coverage of numerical relativity simulations across the large parameter space in which they reside, and numerical relativity simulations are computationally costly and can take weeks to months to finish.
This makes it important to have a method of placing new simulations in ways that use our computational resources optimally while ensuring sufficient coverage of the parameter space.
Accomplishing this requires predicting the impact of a new set of parameters before performing the simulation.
To this effect,  this paper introduces a neural network to predict the mismatch between the gravitational waves of two binary systems.
%The mismatch is a metric frequently used in gravitational wave data analysis to measure the difference between two waveforms and can serve as a way of identifying how important a proposed simulation will be.
Using this network, we then show how we can propose new numerical relativity simulations that will provide the most benefit. 
We also use the network to identify gaps in existing public catalogs and identify degeneracies in the binary black hole parameter space. 
\end{abstract}

\maketitle

\section{Introduction} 

%\note{Current model: mismatch\_22\_flat\_56\_npl\_15\_layers\_2022-07-05\_10-05-15\_0}

%\note{Maybe: Create a surrogate waveform and compute the mismatch there with an actual NR simulation}

As of the end of its third observing run, the \lvk{} has detected 90 \gw{} signals from merging binary systems~\cite{2016PhRvL.116f1102A,PhysRevLett.119.141101,Abbott_2017,PhysRevLett.118.221101,PhysRevLett.116.241103, LIGOScientific:2021djp}.
With the \lvk{'s} fourth observing run, anticipated to begin in 2023, and the fifth observing run a few years later, we expect to observe hundreds of more signals with higher \snr{s}. 
Beyond current detectors, in the coming decades, we also expect to see next generation \gw{} detectors come online, including the space-based \lisa{}~\cite{Robson_2019} and the ground-based \et{}~\cite{Punturo:2010zz, 2011einstein} and \ce{}~\cite{Evans:2021gyd}.
All of these new detectors will have unprecedented sensitivity and are expected to see orders of magnitude more signals than we have to date~\cite{2019arXiv190706482B, 2017arXiv170200786A}.
\lisa{} will be sensitive to lower frequencies than the ground-based detectors, giving us a window into entirely different sources, including supermassive \bh{} binaries which may remain in the frequency band for months to years.

Detecting and interpreting these events requires understanding the expected \gw{} signatures of astrophysical events~\cite{TheLIGOScientific:2017qsa, Abbott:2020uma, Shibata:2017xdx}.
While analytic solutions exist for single \bh{s}, \bbh{s} have no analytic solution for full, nonlinear \gr{}.
Many methods have been developed to find solutions to the \bbh{} problem.
When the \bh{s} are far apart, \pn{} can be used to compute \gw{} waveforms~\cite{Isoyama:2020lls,  Blanchet:2013haa,  Schafer:2018kuf, Futamase:2007zz, Blanchet:2008je, Blanchet:2001ax, Blanchet:2005tk, Faye:2012we, Faye:2014fra}.
In the highly unequal mass ratio regime, self-force approximations can be used to solve \bbh{} systems and output expected waveforms~\cite{Burko:2015sqa, Lackeos:2012de, Hinderer:2008dm, Barack:2018yvs, Poisson:2011nh, Harte:2014wya, Pound:2015tma, Pound:2021qin, vandeMeent:2020xgc, Sperhake:2011ik, LeTiec:2013uey, LeTiec:2011dp, vandeMeent:2016hel, Zimmerman:2016ajr}.

However, in the highly non-linear regime of the merger of comparably massed systems, \nr{} is the most accurate approach~\cite{Loffler:2011ay,  PhysRevD.88.024040}.
\nr{} computationally solves Einstein's equations by performing a 3+1 decomposition, such as in the BSSN formulation~\cite{Baumgarte:1998te},  in order to evolve spacelike surfaces along a timelike path.  
Many \nr{} codes, including those based on the \etk{}, use finite differencing to solve 2nd-order non-linear PDEs~\cite{Schnetter:2003rb, Loffler:2011ay}. 
This enables them to solve for the dynamics of the system and the gravitational radiation emitted~\cite{Schnetter:2003rb, Loffler:2011ay}.
These simulations require significant computational resources; each simulation takes an initial \bbh{} configuration and can run for weeks to months, placing one point in the parameter space of \bbh{s.}
When considering quasicircular, vacuum \bbh{} systems, the parameter space is 7 dimensional, consisting of the spin vectors for each of the initial \bh{s} as well as the ratio of the masses of the initial \bh{s}.
This dimensionality increases when considering eccentric systems or systems with matter. 

Detection pipelines and parameter estimation tools require dense, accurate template banks to detect and interpret the events from the observed detector data~\cite{Capano:2016dsf, alex_nitz_2020_3630601, PhysRevD.90.082004, Usman:2015kfa, Cannon:2011vi, Privitera:2013xza, Messick_2017}.
Due to the high computational cost of \nr{},  in order to produce template banks efficiently and with arbitrary density, analytic models are constructed using information from \nr{}~\cite{Hannam:2013oca, Bohe:2016gbl, Khan:2015jqa, Blackman:2017pcm, Husa:2015iqa, Abbott:2020uma, Taracchini:2013rva}.
In some cases,  \nr{} waveforms are used directly in the analysis~\cite{TheLIGOScientific:2016uux, Lange:2017wki}, to avoid  systematic errors from the models and/or to provide parameter coverage not yet available in models.
In either situation, \nr{} waveforms are an integral part of GW astronomy; therefore, it is crucial to have a dense bank of \nr{} simulations which cover the desired parameter space.
Current public \nr{} catalogs~\cite{Mroue:2013xna,Jani:2016wkt,Healy:2017psd,Boyle:2019kee,Healy:2019jyf,Healy:2020vre} have dense coverage for comparable mass systems, even for those with spins misaligned with the orbital angular momentum.
They are lacking coverage in the precessing space for more unequal massed systems and have only a trace number of simulations where one \bh{} is more than 10 times the mass of the other.
While this coverage has proved acceptable for current \gw{} detector sensitivities, the incredible developments expected in future  \gw{} detectors will require denser, more accurate, and more expansive waveform catalogs~\cite{Ferguson:2020xnm}.  
Achieving the accuracy needed for future \gw{} events requires very high resolutions in current \nr{} codes.
This becomes even more significant for unequal mass and highly spinning \bh{s} as both of these cases require higher resolution to resolve the spacetime around the \bh{s}; this increases the computational cost significantly.

Given the high computational cost and time consuming nature of \nr{} simulations, particularly at high resolution, it is imperative to construct \nr{} catalogs carefully with their cost and use in mind. 
In order to be more strategic about \nr{} simulation placement, we need to understand which simulations will be most beneficial and which regions of parameter space will need to be populated more densely. 
This requires a method of predicting how important a proposed simulation will be before performing it.
Analytic models can be used to this effect by efficiently predicting waveforms for given initial parameters~\cite{Hannam:2013oca, Bohe:2016gbl, Khan:2015jqa, Blackman:2017pcm, Husa:2015iqa, Abbott:2020uma, Taracchini:2013rva}. 
Surrogate models can also be used for this purpose as they interpolate between existing simulations to predict waveforms for given parameters~\cite{Field:2013cfa}.
Using these models to predict the importance of a proposed simulation would require creating a waveform and then comparing it to all existing \nr{} waveforms, a very costly endeavor.
The use of models also adds in potential systematic biases dependent upon the waveform model used. 
Work has also been done to use the uncertainty in models to suggest new simulation parameters~\cite{Doctor:2017csx}

We introduce a more efficient and independent method of estimating the significance of a new simulation using a neural network.
First emerging in 1944 and resurging in the 1980s, neural networks are a type of machine learning that are able to recognize patterns and make predictions based on input values~\cite{McCulloch1943, Hopfield1982, Rosenblatt1958ThePA, werbos1974, LeCun1989}.
The networks consist of a series of layers, each of which may contain many nodes.
Each node functions as a single logistic regression unit, using weights and biases to input many values and output one.
Arranging these nodes in layers allows the network to recognize complex patterns and relationships.
Training the network entails optimizing the weights and biases of the nodes in order for the network to output the desired values for labeled training data.
This process begins by initializing the weights and biases randomly and then propagating the input parameters of the training data to obtain the network output.
This is compared to the desired output for the training data, and backpropagation is used to update the weights and biases accordingly.
This process is repeated until the accuracy reaches a plateau or a specified desired value.  
In recent years, neural networks have been used in the \gw{} and \nr{} fields~\cite{Haegel:2019uop, 2021arXiv210614089Q, Gebhard:2019ldz}.

In this paper, we introduce a neural network to predict the mismatch between two \bbh{} systems given only the initial parameters of the binaries.
The mismatch is frequently used in data analysis to measure the difference between two waveforms, so this allows us to gauge how different a proposed simulation would be from those in existing \nr{} catalogs.
Having a neural network that predicts the mismatch avoids the need to generate waveforms and perform costly mismatch calculations. 
This also allows numerical relativists to use and even train this network on new \nr{} simulations in regions of parameter space that models don't yet cover. 
For instance, while this network is currently trained on quasicircular simulations, it can be extended and retrained on eccentric simulations. 

This paper begins by introducing the match prediction network.
The network, it's training data, and its accuracy are outlined in Section \ref{sec:neural_network}.

Section ~\ref{sec:results} goes through the results of using this network to study and fill in the parameter space.
The primary motivation for the creation of this network is to optimally place \nr{} simulations to fill in catalogs efficiently.
This implementation is described in Section ~\ref{sec:template_placement}.
The network also provides us more insight into the behavior of waveforms throughout the parameter space.
Using this, in Section ~\ref{sec:identifying_gaps}, we note gaps in the current coverage of public \nr{} waveform catalogs.
Section ~\ref{sec:identifying_degeneracies} uses the network to identify degeneracies in parameter space where significantly different parameters can lead to extremely similar waveforms.

The model described in this paper is available for use at \url{https://github.com/deborahferguson/mismatch_prediction}.

\section{Neural Network to Predict Match}\label{sec:neural_network}

In order to identify regions in parameter space that would provide significant scientific benefit if filled, we first need to define a metric to assess how different a resulting waveform would be from existing waveforms.
For this paper, we use the mismatch as our metric~\cite{PhysRevD.53.6749, PhysRevD.57.630}, defined as: 
\begin{equation}
 \mathcal{MM} = 1 - \max_{t,\, \phi} \mathcal O[h_1, h_2]  \equiv  1 - \max_{t,\, \phi} \frac{\langle h_1|h_2 \rangle}{\sqrt{\langle h_1|h_1\rangle \langle h_2|h_2\rangle}},
  \label{eq:overlap}
\end{equation}
where
\begin{equation}
\langle h_1|h_2 \rangle = 2\int_{f_{0}}^\infty\frac{h_1^*h_2 +h_1\,h_2^*}{S_n} df\,.
\end{equation}
$h_1$ and $h_2$ are the frequency domain strain of the gravitational waves, and $S_n$ is the one-sided power spectral density of the detector.
The mismatch is normalized to $0 \leq \mathcal{MM} \leq 1$ with $\mathcal{MM} = 0$ corresponding to two identical waveforms.

With this measure defined, we now need a way to predict it when the waveforms for one or both of the binary systems have not been generated.
This section introduces a neural network that inputs the initial parameters of two quasicircular \bbh{} systems and outputs their predicted mismatch. 

Quasicircular \bbh{} systems are uniquely described by the masses and spins of their component \bh{s}.
All quantities pertaining to the simulations of vacuum \bbh{s} can scale with the total mass, so in \nr{}, the total mass is typically set to 1.
We can then define the symmetric mass ratio, $\eta = m_1 m_2 / (m_1 + m_2)^2$, reducing our parameter space by 1 dimension.
$\eta$ is often used when characterizing the frequency evolution of the inspiral of two \bh{s} and is bounded to $0 \leq \eta \leq 0.25$, where $\eta = 0.25$ corresponds to an equal mass system.
We will use the dimensionless spin vectors, $\bm{a} = \bm{J}/m^2$ where $\bm{J}$ is the angular momentum of the \bh{}.
The magnitude of the dimensionless spin vector is constrained to $0 < a < 1$.

\begin{figure*}
	\centering
	\includegraphics[width = \textwidth]{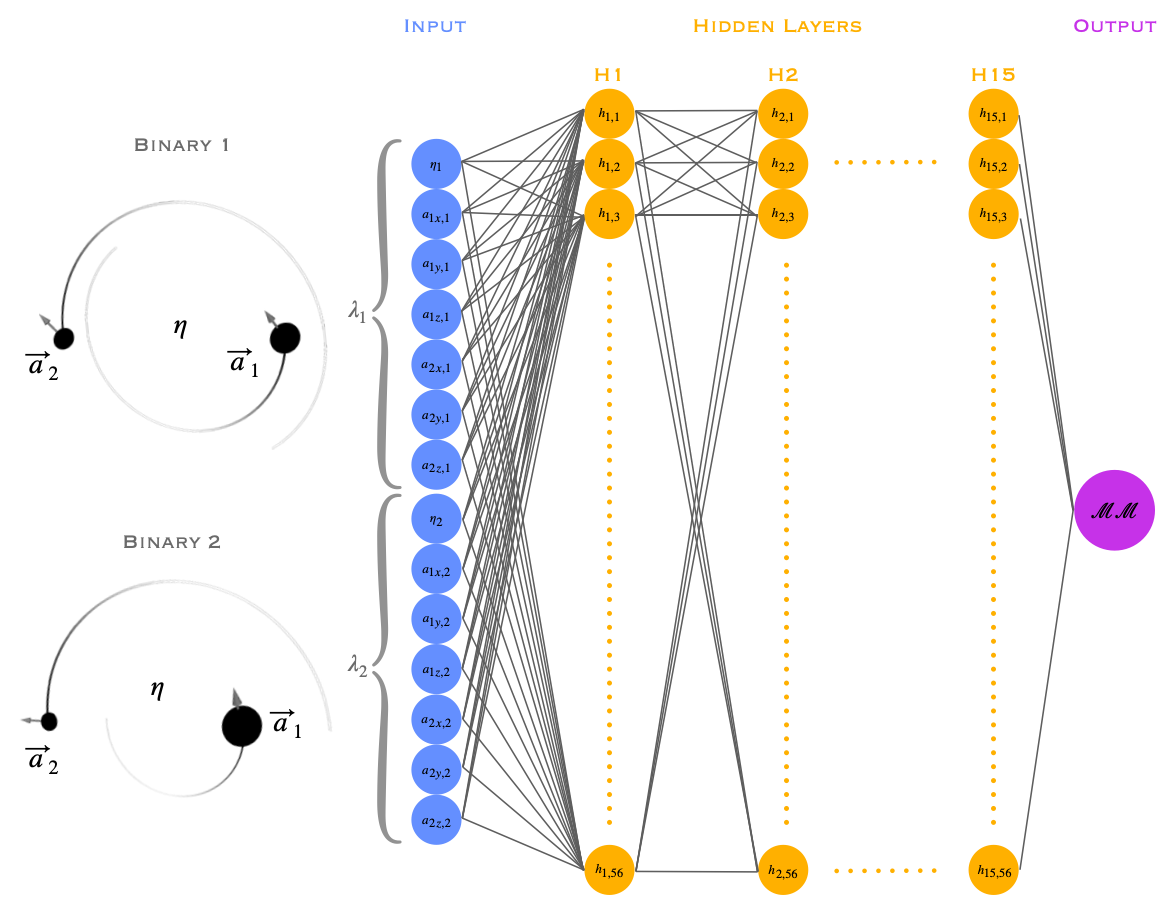}
	\caption{Diagram showing the structure of the neural network including the input parameters ($\eta$, $a_{1x}$, $a_{1y}$, $a_{1z}$, $a_{2x}$, $a_{2y}$, $a_{2z}$ for each of the two binary systems) and the output (the mismatch between the gravitational waves emitted from the two systems).  The network consists of 15 hidden layers, each with 56 nodes.}
	\label{fig:network_design}
\end{figure*}

Let us define $\lambda$ = $\eta$, $\bm{a_1}$, $\bm{a_2}$ to be the initial parameters for a single \bbh{} system.
This network inputs $\lambda_1$ and $\lambda_2$ and outputs the mismatch between the \gw{s} each system would emit. 
This structure (and the architecture of the network as described in Section ~\ref{sec:architecture}) can be seen in Fig. ~\ref{fig:network_design}.

In Section \ref{sec:training_data}, we describe the training data, including the \nr{} waveforms used and the details of the match computation.
Section \ref{sec:architecture} describes the architecture of the neural network, and we discuss the accuracy of the network in Section \ref{sec:accuracy}.

\subsection{Training Data}\label{sec:training_data}

This network is trained on public \nr{} waveforms from the SXS catalog~\cite{Boyle:2019kee}.
The uncertainty in the mismatch due to computational limitations within the training data, including finite resolution and extrapolation, is of order $10^{-4}$.

The training data consists of \bbh{} systems with $0.0826 \leq \eta \leq 0.25$ with spin magnitudes $0 \leq a_1, a_2 \leq 0.9695$ pointing in various directions.
As the symmetric mass ratio decreases, the coverage in the spin parameters becomes less dense.
We discuss how this coverage affects the error in the neural network in Section ~\ref{sec:accuracy}.

For \bbh{} systems where one or both of the \bh{} spins are not aligned or antialigned with the angular momentum, the spins and the orbital plane precess throughout the evolution of the system.
In order to uniquely identify a precessing simulation, the spins must be specified at a given reference point in the simulation; a reference frequency is generally used for this purpose.
To identify the spins at a given reference frequency, we need simulations that reach the specified frequency and contain spin data at that frequency.
Gravitational wave frequency scales with the total mass of the binary, so we define our frequency cutoff in terms of a system with total mass $M_{tot} = 1 M_{\odot}$.
We select $f_{ref} = 1840$ Hz as our reference frequency for spins and as the frequency at which to begin our mismatch calculation.
For the starting frequency of $20$ Hz used by many \lvk{} analyses, this would correspond to a total detector frame mass of $92\, M_\odot$.
This was chosen so as to include 95\% of waveforms in the SXS public catalog.

\begin{figure*}
  \centering
  \includegraphics[width = \textwidth]{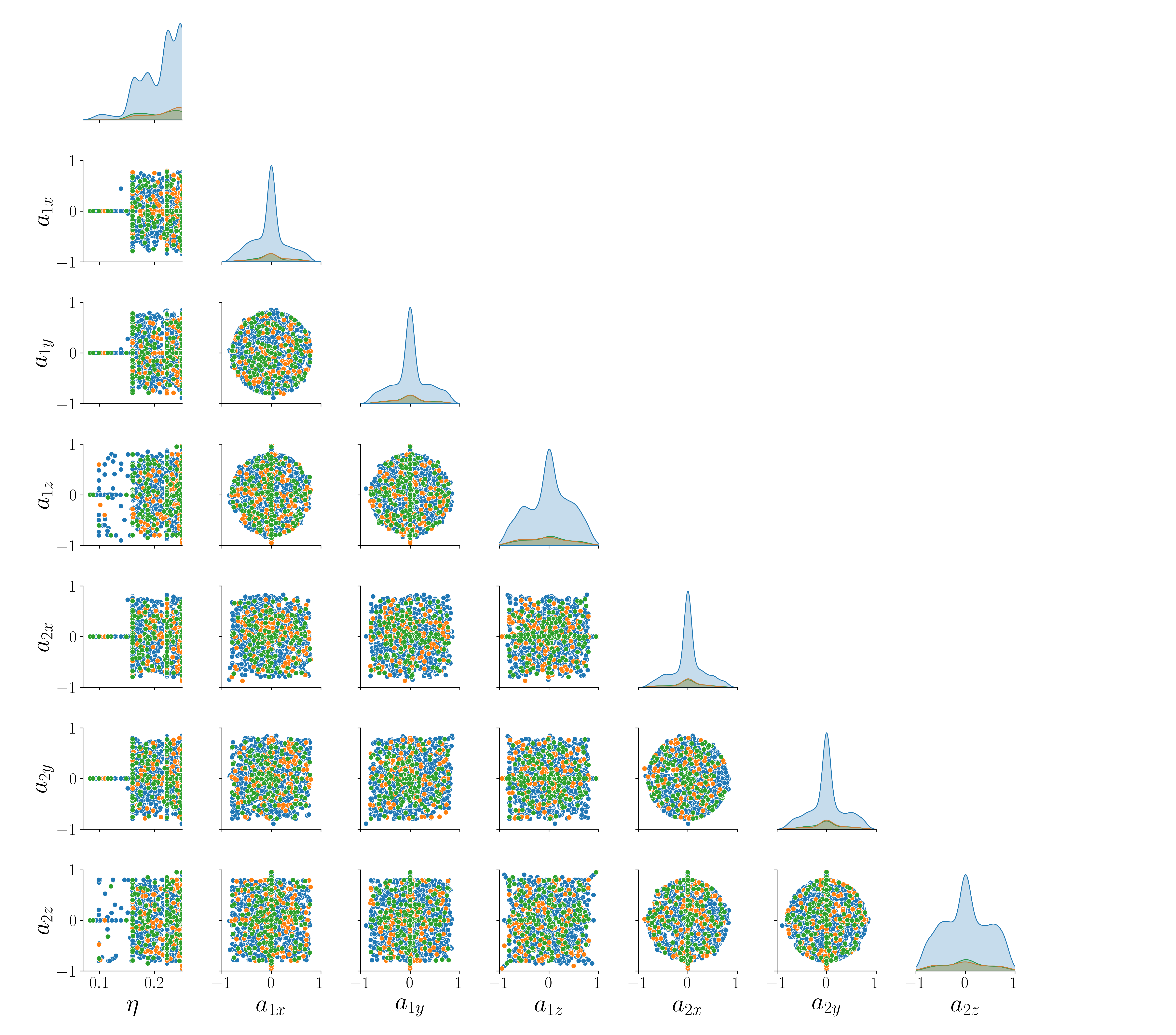}
  \caption{Coverage of the parameter space by \nr{} simulations in training (blue), development (orange), and test sets (green). }
  \label{fig:parameter_space}
\end{figure*}

Selecting only waveforms that have a starting  frequency below $f_{ref}$ and contain spin data at $f_{ref}$ leads to 1885 waveforms.
These are then split into training, development, and test sets with ratios of 0.8/0.1/0.1 respectively. 
Fig. ~\ref{fig:parameter_space} shows the coverage of the parameter space for the \nr{} waveforms used for this analysis.

Using \pycbc{}~\cite{alex_nitz_2020_3630601} and \lalsim{}~\cite{lalsuite}, we compute the mismatch between all pairs of waveforms within each set on a flat noise curve beginning at $f_{ref}$.
This allows us to study the parameter space independently of detector and total mass, though we anticipate training versions of this network on detector noise curves in future work.
For this network, we use only the $\ell = 2, \, m=2$ mode to simplify the mismatch calculation and allow for a phase shift to be used to maximize the overlap.
It also removes the dependence on binary orientation.
While considering only the dominant mode will have an impact on the mismatches for systems where higher modes are significant, it is still informative for exploring the parameter space efficiently, so we reserve the consideration of higher modes for future work.
We define the spin vectors at $f_{ref}$, rotated such that the \bh{s} are separated along the x-axis and the orbital frequency lies along the z-axis.

\subsection{Network Architecture}\label{sec:architecture}
In order to predict the mismatch, we create a fully connected neural network using TensorFlow~\cite{tensorflow2015-whitepaper} and Keras~\cite{chollet2015keras}.
The network consists of 15 hidden layers, each with 56 nodes, as shown in Fig.~\ref{fig:network_design}.
This structure was chosen after exploratory models were created across a grid consisting of the number of hidden layers and nodes per layer.
The development error did not change dramatically for most of the configurations, and the architecture resulting in the lowest development error was chosen. 
Further tuning of the network architecture will likely be a topic of future work.

The network was trained using an Adam optimizer~\cite{2014arXiv1412.6980K} with a learning rate of 0.001 for 100 epochs using a batch size of 16.
In order to add regularization and reduce overfitting, we impose a constrained max norm value of 3.
Each hidden layer uses a ReLU activation function~\cite{2018arXiv180308375A} and the final layer uses a sigmoid function to ensure the resulting mismatch is constrained between 0 and 1.
Predicting mismatches with the network takes very little time since forward propagation through the network can be efficiently parallelized.
The network takes about 3 seconds to load, and then each match requires approximately $3 \times 10^{-5}$ seconds to compute.

\subsection{Network Accuracy}\label{sec:accuracy}
\begin{figure*}[tp]
	\centering
	\begin{subfigure}[t]{0.54\textwidth}
   		\includegraphics[width = \textwidth]{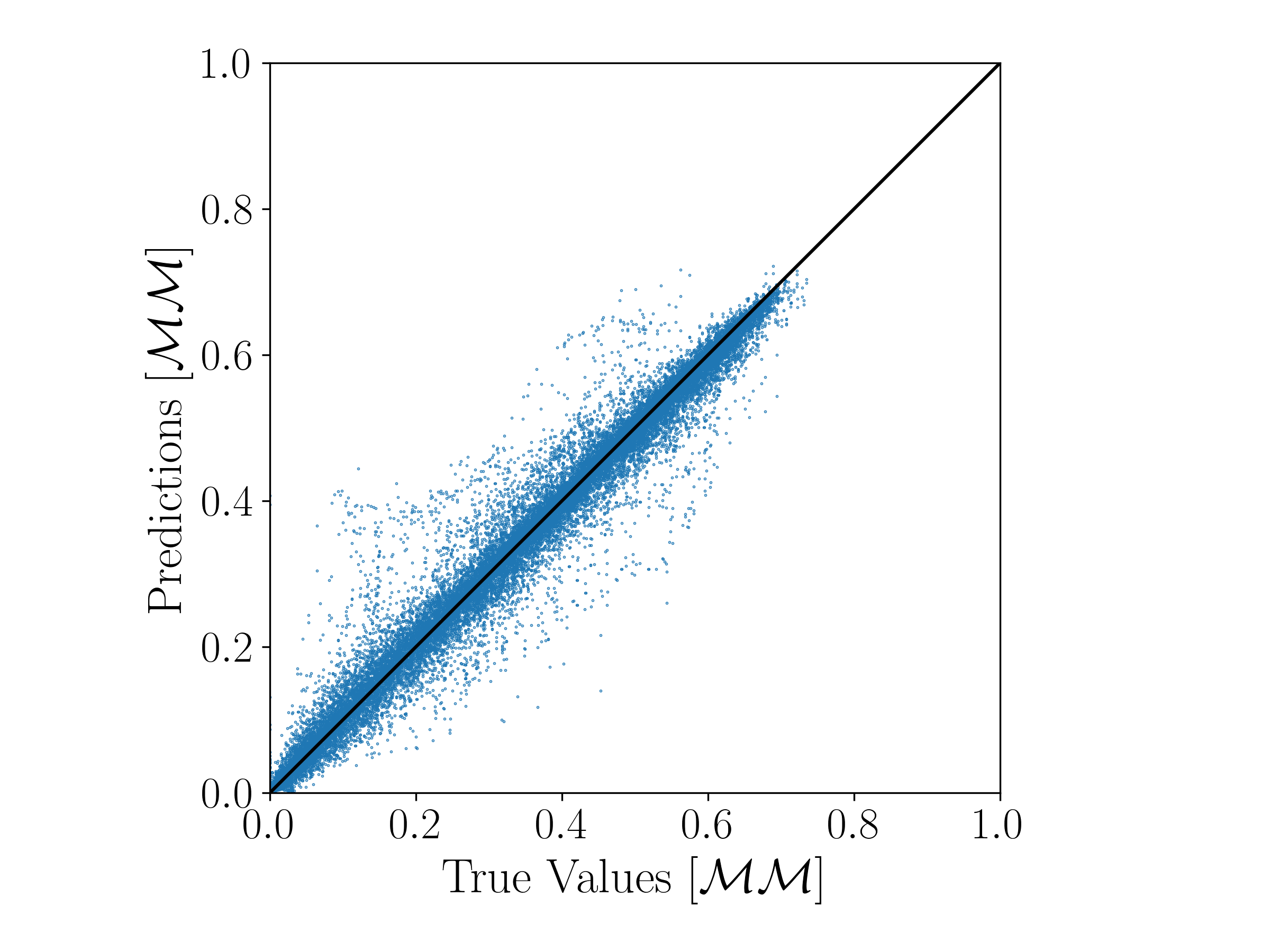}
   		\caption{The mismatches predicted by the network versus the mismatches computed from the waveforms in the test set.}
   		\label{fig:error_line}
	\end{subfigure}
	~
	\begin{subfigure}[t]{0.4\textwidth}
   		\includegraphics[width = \textwidth]{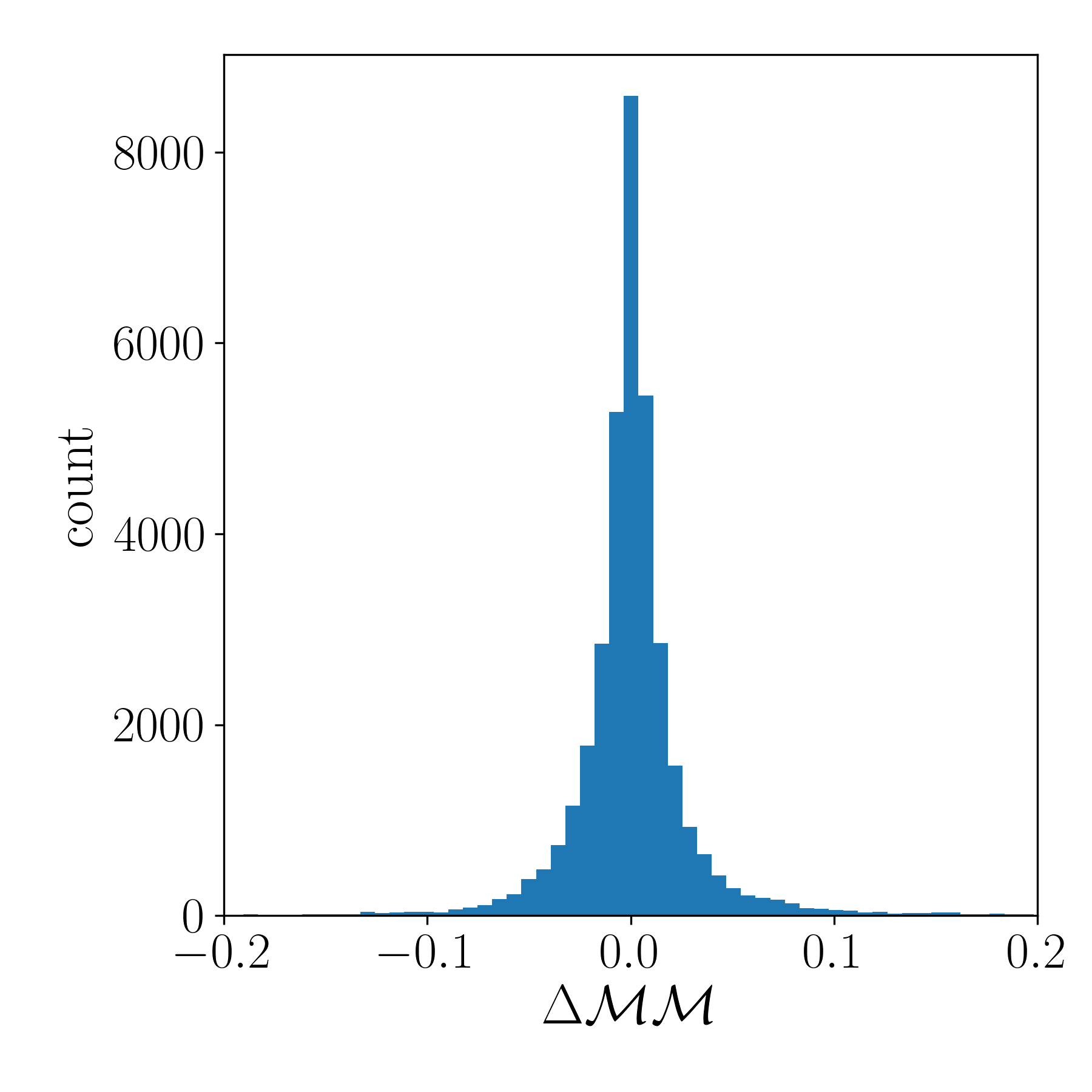}
   		\caption{The distribution of the error in the predicted mismatches for waveforms in the test set.}
   		\label{fig:error_hist}
	\end{subfigure}	
	
	\caption{The error in the predicted mismatch for the waveforms in the test set.}
	\label{fig:error}
\end{figure*}

To compute the accuracy of the network on data it has not seen, we take the simulations reserved for the development set and compute the mismatches between all possible simulation pairs.
These mismatches are compared to the predicted mismatches and the mean absolute error is computed.
This is also done for the test set.
After training the network for 100 epochs, it has a training error of 0.00967, a development error of 0.0157, and a test error of 0.018.
Therefore, our uncertainty in our mismatch prediction is 0.018.
%This is comparable to the match between two equal mass systems, each with one \bh{} spinning aligned with the orbital angular momentum that have a difference in spin magnitude of $\Delta a = \pm 0.1$.
Fig.~\ref{fig:error_line} shows the predicted mismatch versus the computed mismatch for the test set.
In the absence of all error, the points would fall on the $y=x$ line.
Fig. ~\ref{fig:error_hist} shows the distribution of the error in the predicted mismatch for the test set.

This network inputs $\lambda_1$ and $\lambda_2$ resulting in 14 input parameters.
The mismatch is order independent so the order of $\lambda_1$ and $\lambda_2$ in the input layer should not affect the predicted mismatch.
After training the network,  when swapping the order of $\lambda_1$ and $\lambda_2$, we find that the average difference between the predicted mismatches is 0.016, which is less than the network's uncertainty. 

\begin{figure}
	\centering
	\includegraphics[width = 0.5\textwidth]{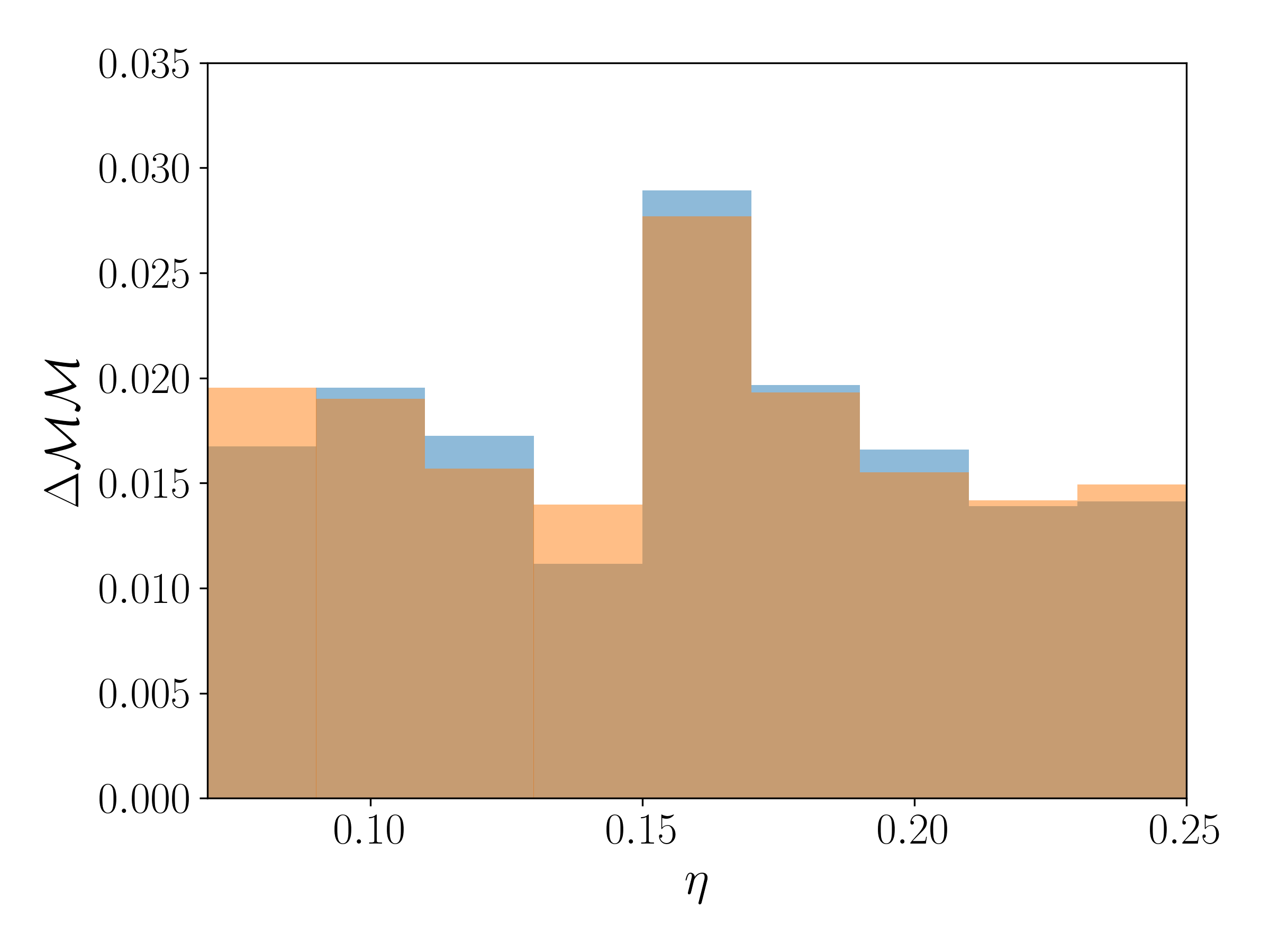}
	\caption{The mean absolute error in the mismatch for the test set binned by the symmetric mass ratio. The blue represents the first system and the orange represents the second system.}
	\label{fig:mass_ratio}
\end{figure}

\begin{figure}
	\centering
	\includegraphics[width = 0.5\textwidth]{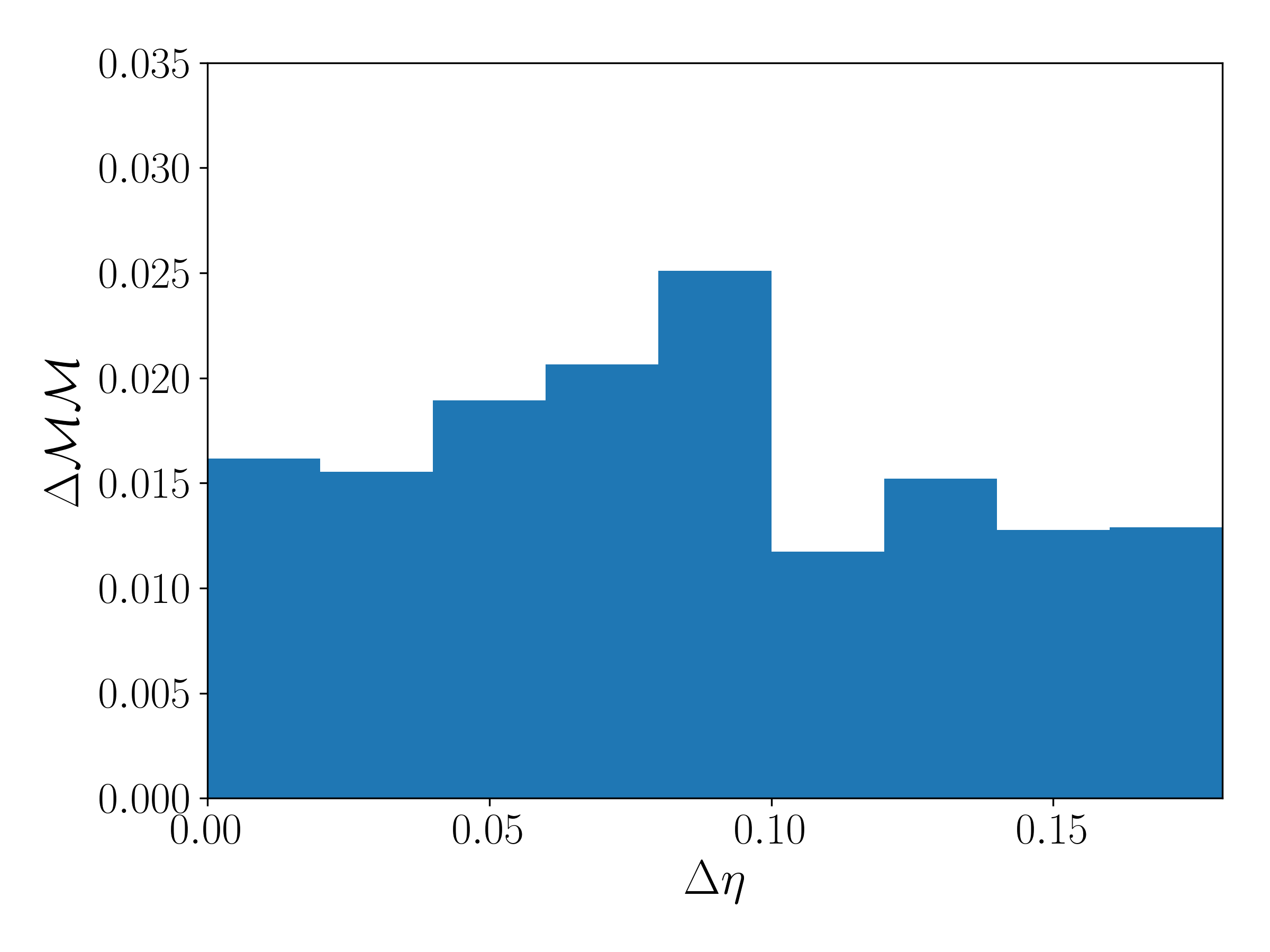}
	\caption{The mean absolute error in the mismatch for the test set binned by the difference in the symmetric mass ratio between the two systems.}
	\label{fig:delta_mass_ratio}
\end{figure}

\begin{figure*}
	\centering
	\includegraphics[width = \textwidth]{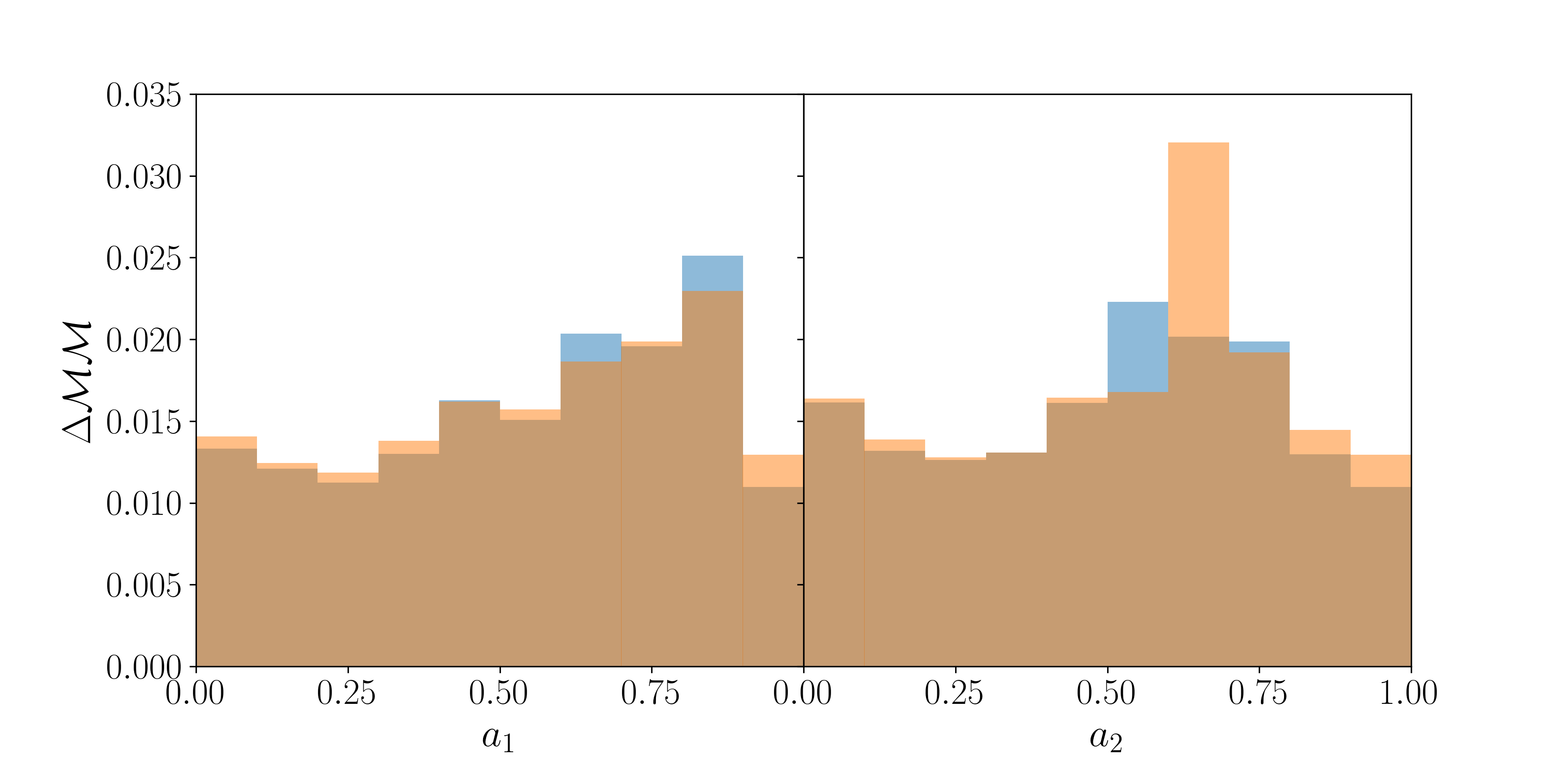}
	\caption{The mean absolute error in the mismatch for the test set binned by the magnitudes of the spins of the primary (left) and secondary (right) black holes.  The blue represents the first system and the orange represents the second system.}
	\label{fig:spin_magnitudes}
\end{figure*}

\begin{figure*}
	\centering
	\includegraphics[width = \textwidth]{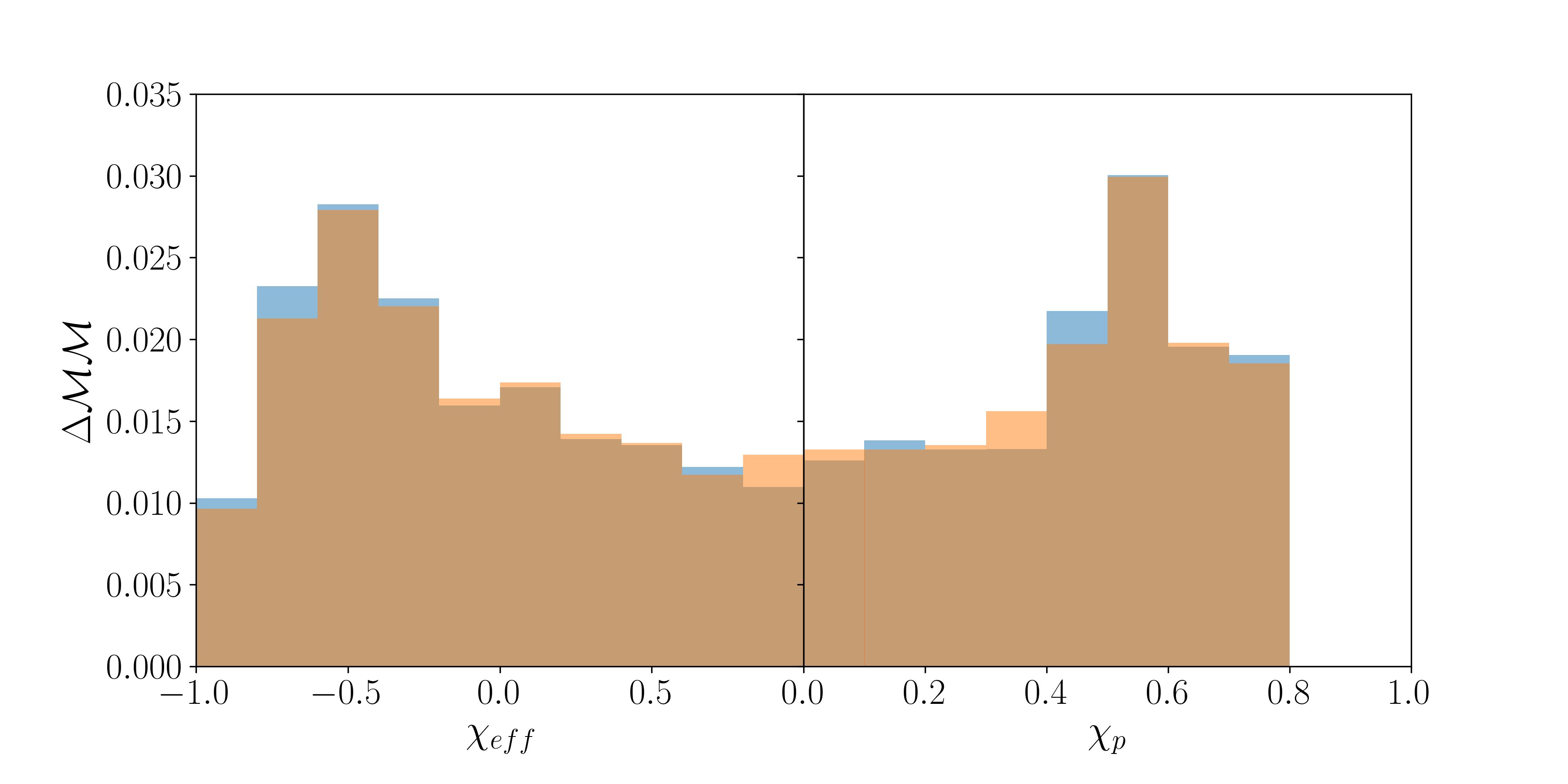}
	\caption{The mean absolute error in the mismatch for the test set binned by $\chi_{eff}$ (left) and $\chi_{p}$ (right). The blue represents the first system and the orange represents the second system.}
	\label{fig:chi}
\end{figure*}

\begin{figure*}
	\centering
	\begin{subfigure}[b]{\textwidth}
   		\includegraphics[width = \textwidth]{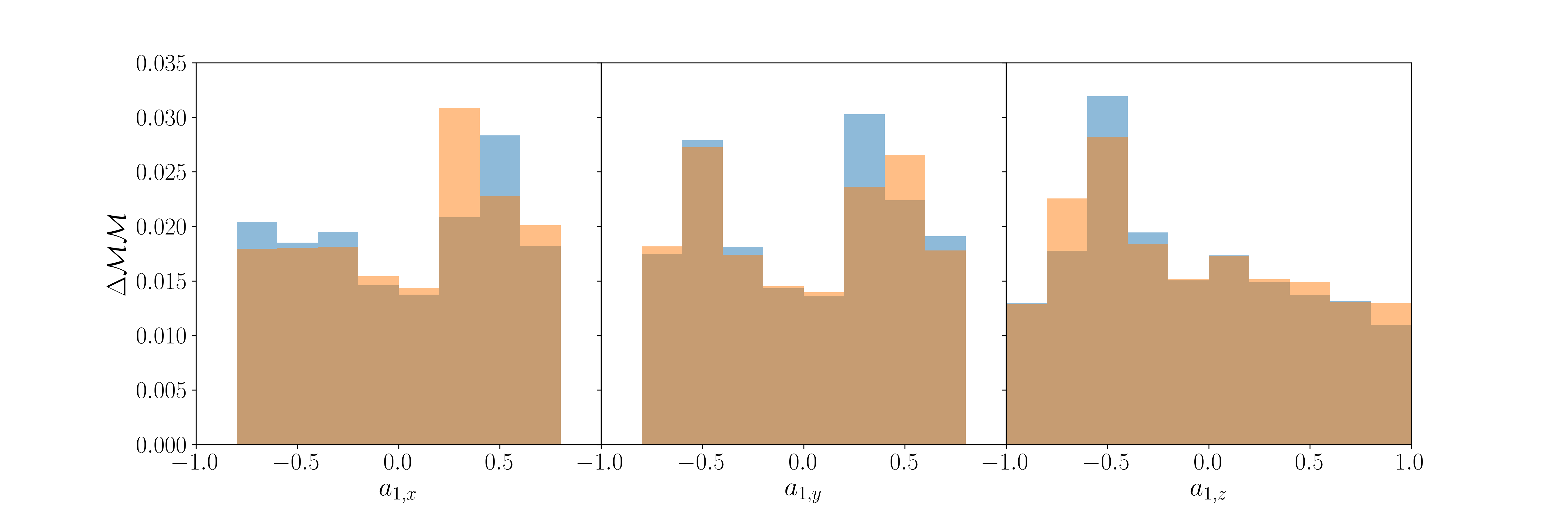}
   		\caption{primary dimensionless spin components}
   		\label{fig:primary_spin_components} 
	\end{subfigure}

	\begin{subfigure}[b]{\textwidth}
   		\includegraphics[width = \textwidth]{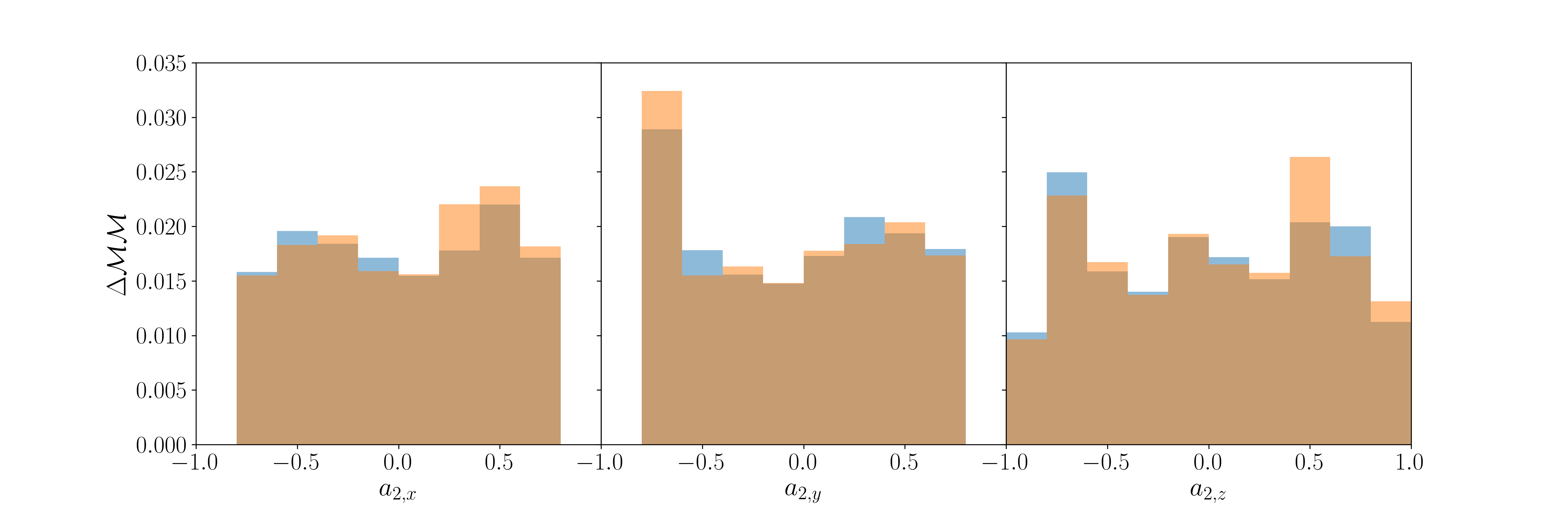}
   		\caption{secondary dimensionless spin components}
   		\label{fig:secondary_spin_components} 
	\end{subfigure}	
	
	\caption{The mean absolute error in the mismatch for the test set binned by the components of the spin. The blue represents the first system and the orange represents the second system.}
	\label{fig:spin_components}
\end{figure*}

Since the training data is not evenly dispersed throughout the entire parameter space, it is expected that the error in the network is not constant in all regions of parameter space.
Figures ~\ref{fig:mass_ratio} - ~\ref{fig:spin_components} show the mean absolute error binned by the various input parameters to the network. 
Since the network inputs the parameters for both binaries,  several of the Figures ~\ref{fig:mass_ratio} - ~\ref{fig:spin_components} overlay the error distributions associated with each of the two systems,  with the first system in blue and the second in orange. 

Fig. ~\ref{fig:mass_ratio} shows the mean absolute error of the model binned by the symmetric mass ratio with bin widths of 0.02. 
The average error goes up with decreasing $\eta$, as is expected due to the scarcity of data points at low $\eta$.
There is also an interesting spike in the error around $\eta=0.16$. 
Referring back to Fig.~\ref{fig:parameter_space}, it is apparent that the density of points drops off by $\eta < 0.16$, so it is expected that the error would be high there, though the reason for the specific peak is not obvious.

In Fig.~\ref{fig:delta_mass_ratio}, we see the mean absolute error binned by the difference in the symmetric mass ratio between the two binary systems.
A peak is apparent around $\Delta \eta = 0.09$
This is consistent with the peak at $\eta = 0.16$ since most simulations fall at $\eta = 0.25$.
For cases that have a binary with $\eta = 0.16$,  we will have many points with high error at $\Delta \eta = 0.09$.

In Fig.~\ref{fig:spin_magnitudes}, the data is binned by the spin magnitudes for both the primary and secondary black holes with bins of width 0.1. 
For the primary black hole, the error grows with increasing spin magnitude, peaking at $a_1 \approx 0.85$.
The error binned by the secondary spin magnitude shows a wide peak around $a_1 \approx 0.7$.
In both cases, there is a minimum error around $a_{1,2} \approx 0.25$.

The first panel of Fig.~\ref{fig:chi} bins the data by $\chi_{eff}$, a quantity used to represent the effective spin of the binary aligned with the orbital angular momentum~\cite{PhysRevD.64.124013}:
\begin{equation}
\chi_{eff} = \frac{m_1 a_1 cos(\theta_1) + m_2 a_2 cos(\theta_2)}{m_1 + m_2} 
\end{equation}
where $\theta_1$ and $\theta_2$ are the angles of the \bh{} spin vectors with the orbital angular momentum.
$\chi_{eff}$ ranges from -1 to 1, with $\chi_{eff} = 1$ occurring when both \bh{s} are maximally spinning and aligned with the orbital angular momentum and $\chi_{eff} = -1$ occurring when both \bh{s} are maximally spinning and anti-aligned with the orbital angular momentum.

The second panel of Fig.~\ref{fig:chi} bins the data by $\chi_p$, a parameter used to describe the precession of a binary~\cite{Racine:2008qv, Schmidt:2014iyl, Schmidt:2012rh, Hannam:2013oca}:
\begin{equation}
\chi_{p} = max\left(a_1 sin(\theta_1), \frac{4 m_2 ^ 2 + 3 m_1 m_2}{4 m_1 ^2 + 3 m_1 m_2} a_2 \theta_2 \right) \, .
\end{equation}
When the spins of the component \bh{s} are not aligned with the orbital angular momentum, they drive a precession of the orbital plane; $\chi_p$ quantifies this effect.
Ranging from 0 to 1, a value of $\chi_p = 0$ signifies no precession, i.e.  any spin is aligned with the orbital angular momentum and the orbital plane will not precess, and a value of $\chi_p = 1$ denotes maximal precession. 

Other than the dip at $\chi_{eff} \approx -1$, the error seems to be higher at strongly negative $\chi_{eff}$ and lower at positive $\chi_{eff}$.
This may be explained by the fewer data points at negative $\chi_{eff}$ than positive.
The error seems to generally increase as $\chi_p$ increases, though it seems to peak at around $\chi_p \approx 0.6$.
This is expected due to the relative scarcity of precessing training points at high $\chi_p$, particularly as $\eta$ decreases.

Fig.~\ref{fig:spin_components} shows the error when the data is binned by each of the components of the primary (a) and secondary (b) spin. 
For both the primary and secondary black hole, the error seems to grow as the x- or y-component of the spin moves away from 0.
These results are anticipated and consistent with the errors when binned by $\chi_p$.
For the primary black hole, the error seems to grow at negative $a_z$, consistent with the results for $\chi_{eff}$.
The trends seem more pronounced for the primary black hole than the secondary black hole, which is expected given that the spin of the primary black hole is known to play a larger role in the shape of the waveform.

Looking specifically at outlier points with error greater than 0.2, we find that the average $\chi_eff$ is strongly negative and $\chi_p$ is higher than in the dataset as a whole.
The trends in the error appear consistent with expectations due to the density of training points throughout the parameter space.
As we perform new simulations and add them to the training data, we expect these errors to decrease.

\section{Results}\label{sec:results}
This network enables us to explore the parameter space more efficiently and fill it in more effectively.
With the network in hand to predict how significant a simulation will be, we can create an automated tool to suggest the next simulation that should be performed. 
This helps us optimally use the computational resources and time available to us.
With the high computational cost of \nr{} simulations, this will be particularly important as we prepare dense catalogs for future \gw{} observations.
Section ~\ref{sec:template_placement} introduces such a tool.

Similarly, by predicting the mismatch between points covered by existing catalogs and points not yet populated, we can note gaps in the parameter space that haven't been filled to a certain threshold in mismatch.
There are regions of parameter space known to be insufficiently filled, but there are other regions whose statuses are less clear.
Identifying gaps in coverage will be important for assessing and improving the accuracy of models.
Section ~\ref{sec:identifying_gaps} shows such gaps in the current public waveform catalogs.

Both of the above scenarios have involved using the network to predict the mismatch between unpopulated and populated points in parameter space.
We can also use the network simply to explore the mismatch behavior across the parameter space.
By doing this, we can note degeneracies, points in parameter space that are significantly separated in terms of spins or mass ratio and yet have low mismatch with one another.
Section ~\ref{sec:identifying_degeneracies} utilizes the network to explore such degeneracies. 

\subsection{Template Placement}\label{sec:template_placement}

There are several approaches one could take to identify the next simulation to perform depending on one's needs.
In this section, we introduce a tool to use this new network to propose parameters for a simulation that would maximize the minimum mismatch between the new simulation and existing simulations.
In other words, it maximizes the distance between the new simulation and its nearest neighbor (with distance characterized by the mismatch).
This will efficiently cover the edges of the desired parameter space and populate gaps within it.

The current parameter space coverage does not lead to a very smooth function to maximize; we have a seven dimensional parameter space with many local maxima, so the suggestion for the next simulation depends greatly upon the initial guess fed to the maximizing function.
To find a global maximum in the presence of local maxima, we utilize the scipy implementation of basin hopping~\cite{doi:10.1021/jp970984n}.
Developed to minimize the energy of a cluster of atoms, basin hopping is a method of identifying the global extremum when many local extrema may be present.
The method cycles through two steps, a local minimization, and then a perturbation.
Given a starting point, it performs a local minimization, finding the nearest local extremum.
It then perturbs the coordinates and performs another local minimization beginning at the new coordinates.
Depending on the height of the newly found extremum, the new coordinates are either accepted or rejected.
Another perturbation is then performed from either the new or old coordinates. 
This cycle continues for a given number of iterations. 

Without doing a full brute force analysis, we can't be certain if a global maximum has been found.
Instead, we can apply the same process from multiple different starting points and test whether they converge to the same maximum.
We find that while they don't approach the exact same parameters, they do all approach maxima of nearly the same height with mismatches of $\approx 0.623 \pm 0.016$.
The difference in the peaks is within the uncertainty of the network and implies the presence of multiple, very comparable local maxima, any of which could be deemed the global maximum.

\begin{figure*}
	\centering
	\includegraphics[width = \textwidth]{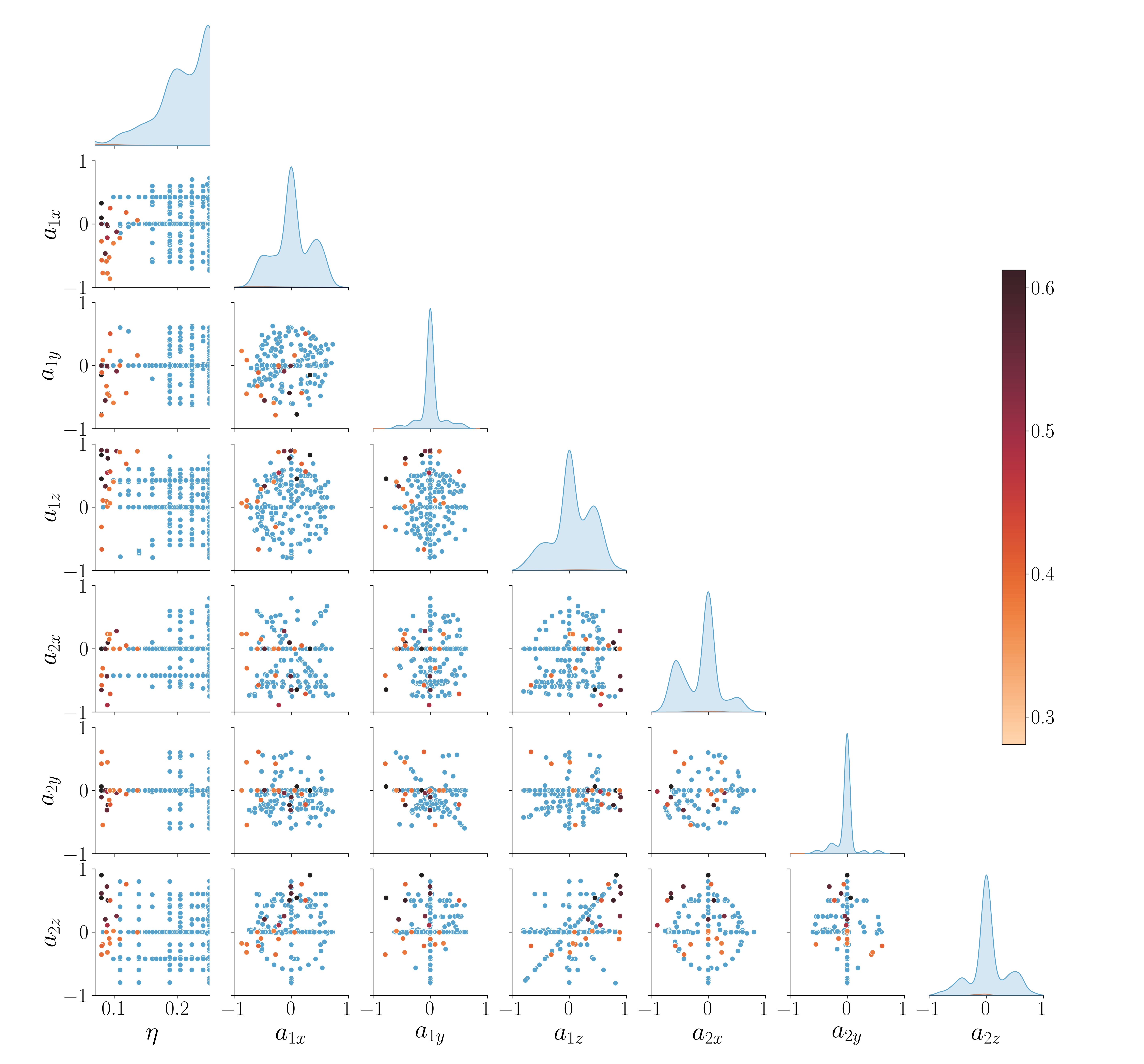}
	\caption{Suggested simulations, selected so as to maximize their mismatch with existing simulations. Blue points are existing simulations from the MAYA catalog, and red points are suggested new simulations. The shade denotes a point's mismatch with its nearest neighbor at the time it is suggested,  with darker being a higher mismatch.}
	\label{fig:suggested_simulations}
\end{figure*}

Now that we're confident that the basin hopping optimization method is finding the parameters that lead to a maximal mismatch, we can apply it multiple times to suggest future simulations.
In this paper, we use a greedy approach, proposing simulations one at a time.
After a simulation is suggested, we add it as an existing data point and then perform another optimization to search for the next best point.
Fig.~\ref{fig:suggested_simulations} shows the results of doing this 20 times, restricting $\eta \geq 0.0826$.

The parameters output by the optimization algorithm are generally in regions where we would anticipate requiring more points. 
All the suggested points have a symmetric mass ratio of $\eta \lesssim 0.16$ and many have higher $x$ and $y$ spin components to fill in gaps in the precessing space.
The mismatch between the first suggested simulation and its nearest neighbor was $\sim 0.6$ while the mismatch between the 20th proposed simulation and its nearest neighbor was $\sim 0.3$.
With only 20 simulations, we have reduced the largest mismatch resulting from a gap in parameter space by a factor of 2. 

While the parameters are in areas that aren't entirely surprising, they improve the parameter space coverage much more efficiently than maximizing the separation of the parameters would. 
To assess this, we used the same optimization technique to optimize for the maximum parameter separation rather than mismatch.
As a metric for the separation in the parameter space, we compute the magnitude of the separation along each of the seven axes normalized by the considered range.
It does begin to fill in the parameter space, but after 20 simulations, there are still regions of parameter space with a mismatch of 0.6 with their nearest neighbor.

\subsection{Identifying Gaps}\label{sec:identifying_gaps}

\begin{figure*}
	\centering
	\includegraphics[width = \textwidth]{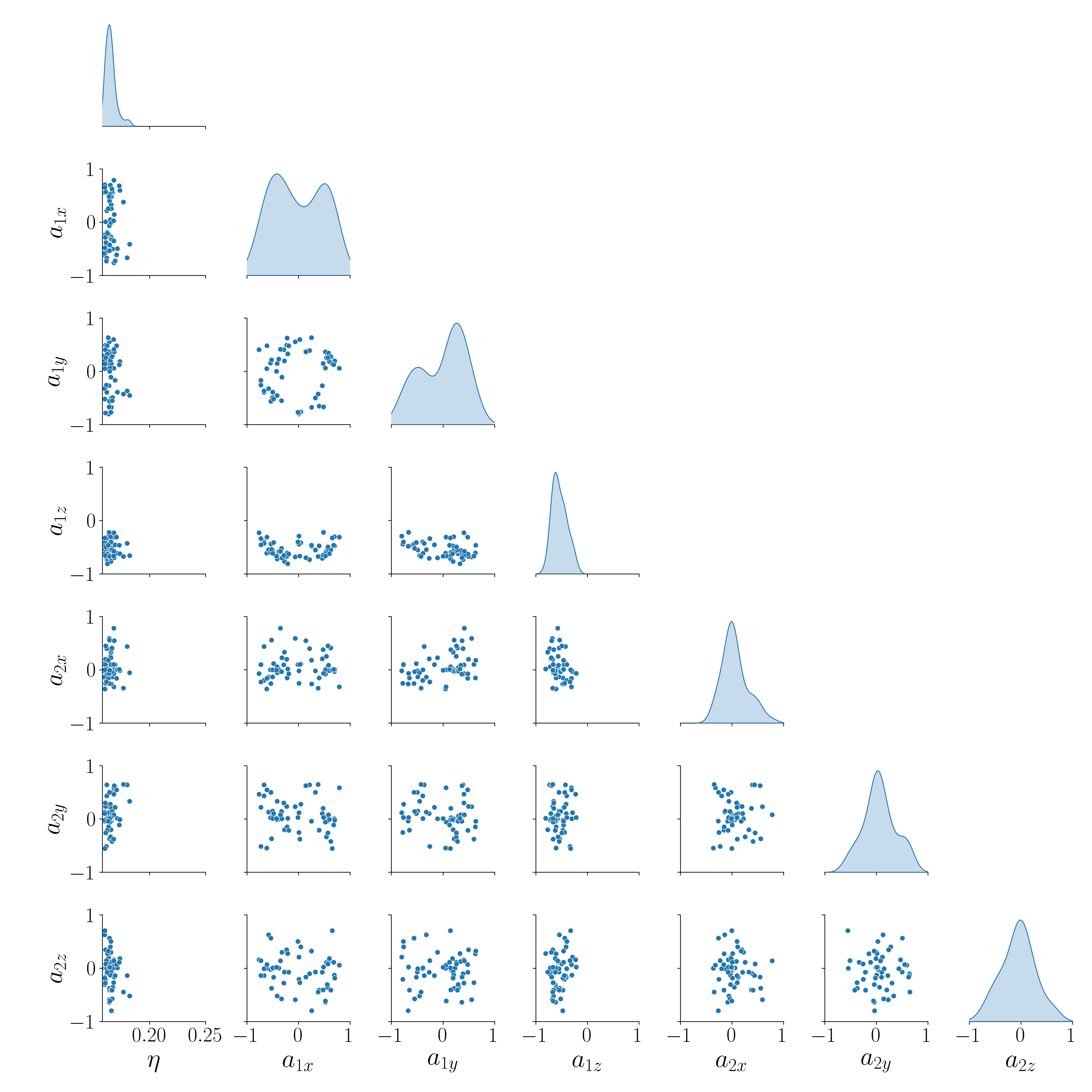}
	\caption{Randomly sampled points whose mismatch with their nearest neighbor in existing public \nr{} waveform catalogs is greater than 0.1.}
	\label{fig:gaps}
\end{figure*}

There are several public \nr{} catalogs including those from SXS~\cite{Boyle:2019kee}, RIT~\cite{Healy:2017psd, Healy:2019jyf}, and Maya~\cite{Jani:2016wkt}. 
Combining these three catalogs, there are some regions of parameter space that appear to have dense coverage and others that appear rather sparse.
This section reports on a brute force analysis to identify gaps within the coverage of these three public catalogs.
We randomly sample 100,000 points across the seven dimensional parameter space and, for each point, find their lowest mismatch with existing simulations. 
Those points whose mismatch with their nearest neighbor is greater than 0.1 are shown in Fig. ~\ref{fig:gaps}.
The uncertainty of 0.018 in the mismatch prediction contributes some uncertainty to these points, but it is clear that any point included in this figure still represents a significant gap.

We only show simulations with $\eta \geq 0.16$ in this figure since it is well known that the parameter space below $\eta=0.16$ is sparsely populated. 
By including only those points with $\eta \geq 0.16$, we can identify gaps in what appears to be a well populated region of parameter space.
Of the 53,019 points in this region, 57 have a mismatch with their nearest neighbor greater than 0.1.
The \nr{} catalogs have dense coverage for $\eta \geq 0.19$, but in the region $0.16 \leq \eta \leq 0.19$,  more simulations are needed.
In particular, in terms of the primary \bh{}, more simulations are needed with anti-aligned or precessing spins.
In terms of the secondary \bh{}, all spin configurations are still needed. 

To understand the impact of these gaps on the training of analytic models, we assess the performance of IMRPhenomPv3HM~\cite{2020PhRvD.101b4056K} at two points in the precessing space, one in a region densely populated by \nr{} points (mismatch of 0.997 with the nearest catalog point) and one in a gap (mismatch of 0.839 with nearest catalog point).
We find that the mismatch between the model waveform and the \nr{} waveform is about an order of magnitude higher for the system placed within a gap.

\subsection{Identifying Degeneracies}\label{sec:identifying_degeneracies}
It is clear that changing the parameters of the initial \bh{s} affects the morphology of the gravitational waves their coalescence emits.
Many of the patterns of these effects are well known, and in some cases, changing different parameters can lead to similar effects on the waveform. 
It may even be possible that two binary systems with very different initial parameters may lead to extremely similar gravitational waves.
In such a situation,  analysis pipelines may not be able to distinguish between the potential sources.
This can complicate the population studies performed on \gw{} catalogs~\cite{LIGOScientific:2016vpg, LIGOScientific:2020kqk, Zevin:2020gbd}.
In some cases, degeneracies may only appear in certain modes and can be broken when analyzing other modes. 
If we identify where such degeneracies occur, we can study the systems further to identify any degeneracy breaking properties.
%This may motivate the inclusion of more higher modes in models or the direct use of \nr{} simulations in analysis and follow-up.
%For example, for a given event, in addition to performing follow-up \nr{} simulations at the maximum likelihood points, we could perform simulations at expected degenerate points to search for differences and provide more confidence to the parameters claimed. 

\begin{figure*}
	\centering
	\includegraphics[width = \textwidth]{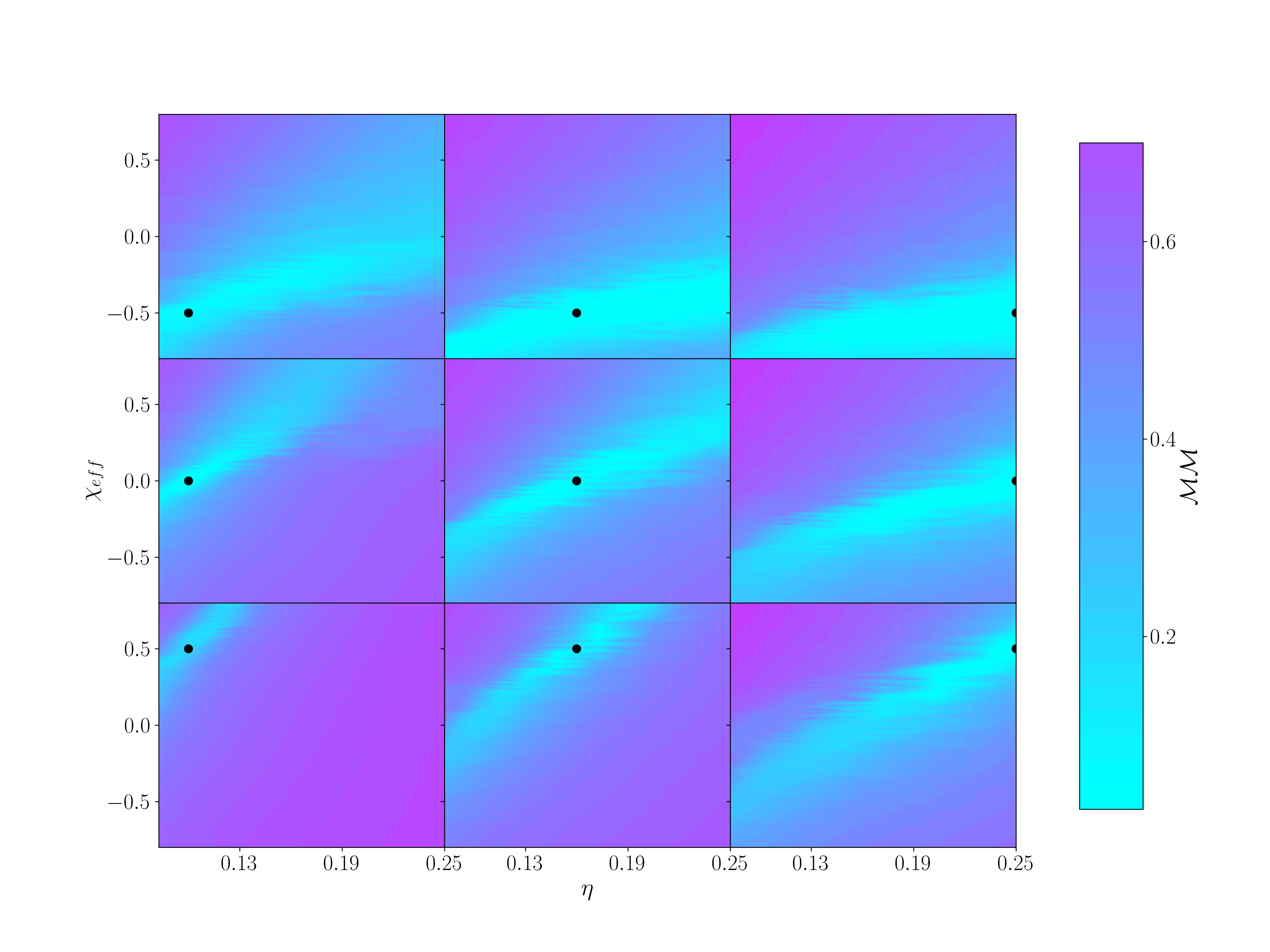}
	\caption{Mismatch between a reference simulation and other simulations throughout the parameter space.  All simulations considered here have \bh{} spins aligned with the orbital angular momentum. The black dot in each panel denotes the reference simulation. From left to right, the reference simulations have symmetric mass ratios of $\eta= 0.1, 0.16, 0.25$, and from top to bottom, the reference simulations have $\chi_{eff}=-0.5, 0, 0.5$.}
	\label{fig:degeneracies}
\end{figure*}

The network presented in this paper can be used to identify such degenerate regions of parameter space. 
Since a seven dimensional parameter space is challenging to show visually, we consider the aligned space here as an example, meaning we are only considering systems with spin vectors aligned with the orbital angular momentum of the binary ($a_{1x}=a_{1y}=a_{2x}=a_{2y}=0$).
This also equates to systems where $\chi_p=0$.
Figure ~\ref{fig:degeneracies} shows the mismatch between a reference binary and other binaries throughout the parameter space, plotted as $\chi_{eff}$ vs $\eta$.
This shows the distribution of the predicted mismatch across the aligned spin parameter space when one of the binaries is held fixed.
Each panel uses a different reference simulation, denoted by the black dot.

Both low $\eta$ and high $\chi_{eff}$ cause the binary system to take longer to inspiral when compared to equal mass, nonspinning binaries; this similarity can be seen in Figure ~\ref{fig:degeneracies} where low $\eta$,  low spinning simulations have a very low mismatch with high $\eta$, highly spinning simulations. 
We can see that even binaries with symmetric mass ratios as low as $\eta \sim 0.08$ can be confused with a highly spinning, equal mass binary. 
This could have significant implications for population studies, as highly spinning events and highly unequal mass events could be confused.
Since this network is trained only on the $\ell = 2, \, m=2$ mode, it is possible that this degeneracy will be broken in higher modes; this motivates further study of these systems.

\section{Conclusions}
\nr{} simulations form an important basis for the detection and characterization of \gw{} signals.
Whether they are used directly with data analysis or to train analytic models, it is crucial to have a sufficiently dense template bank of \nr{} simulations.
Given the high computational cost of \nr{} simulations, it's important to select initial parameters with care.
In this paper, we have introduced a way to place new \nr{} simulations using a neural network trained to predict the mismatch between the waveforms of two binary systems.

The network has an average uncertainty of 0.018 in the predicted mismatch.
Due to the high density of training points at high $\eta$,  the error is less at high $\eta$ and increases as $\eta$ decreases, particularly for precessing binaries. 
Even for the most challenging regions of parameter space, the uncertainty stays below $\sim 0.03$.

With this network, we were able to optimally select initial parameters so as to maximize the mismatch of a new simulation with its nearest neighbor.
With only 20 simulations, we decreased the maximal mismatch in unpopulated regions by a factor of 2.
As we prepare for the next generation of \gw{} detectors, it becomes crucial to have a densely populated template bank of highly accurate waveforms.
This method will aid us in creating such a template bank, ensuring every simulation performed is chosen to maximize our coverage.

We also used the mismatch predicting network to identify gaps in the coverage of current public \nr{} waveform catalogs.
It is clear that more simulations are needed at $\eta < 0.16$, but our analysis also shows that in the region of $0.16 \leq \eta \leq 0.19$, there are significant coverage gaps.
This is particularly true for binaries with anti-aligned or precessing spins. 

The network also provided us a method to explore the parameter space, seeing how the mismatch changes with each parameter and identifying degeneracies.
We demonstrated how highly unequal mass, low spinning simulations can be confused with comparable mass, highly spinning simulations, encouraging more study into how these degeneracies may be broken.

As we prepare for the next generation of \gw{} detectors, we need to have a strong understanding of \bbh{} systems within \gr{.}
The network presented in this paper to predict the mismatch between binary systems provides a unique and powerful approach to studying the \bbh{} parameter space.
The template placement tool utilizes this network to optimally propose simulations that will efficiently expand our \nr{} catalogs,  greatly improving our readiness for the coming improvements to \gw{} detectors.

\paragraph*{\textbf{Acknowledgements}}
This work is supported by grants NASA 80NSSC21K0900 and NSF PHY 2207780. 
Computing resources were provided by XSEDE PHY120016 and TACC PHY20039.
The author thanks Deirdre Shoemaker for support and helpful discussions.
The author also thanks Harald Pfeiffer for useful feedback.
This work was done as part of the Weinberg Institute, with the identifier UTWI-20-2022.

%\bibliography{refs,nlst_wp}
\bibliography{ms}
\end{document}